\documentclass[12pt,journal]{IEEEtran}
\usepackage{graphicx}
\usepackage{orcidlink}
\usepackage{float}
\usepackage{bm}
\usepackage{subfigure}
\usepackage{amsmath,amsfonts}
\usepackage{mathtools}
\usepackage{algorithm}
\usepackage{algorithmic}
\usepackage[normalem]{ulem}
\usepackage[tableposition=top]{caption}
\usepackage{url}
\usepackage{booktabs}

\usepackage{array}
\usepackage{cite}
\usepackage{color}
\usepackage{cleveref}
\usepackage{geometry}
\usepackage{amssymb}
\usepackage{cuted}
\usepackage{lipsum} 
\usepackage{multicol}
\usepackage{amsmath,amssymb,amsthm,bm}

\newgeometry{left=1.4cm,right=1.4cm,bottom=1.9cm,top=1.5cm}

\begin{document}


\title{Virtualizing the Senses: Enabling High-Precision ISAC on Commercial Cellular Infrastructure}

\author{
Henglin Pu and
Husheng Li, \IEEEmembership{Senior Member, IEEE}

\thanks{Henglin Pu and Husheng Li are with the Elmore Family School of Electrical and Computer Engineering, Purdue University, West Lafayette, Indiana, USA-47907 (email:
 pu36@purdue.edu, husheng@purdue.edu).}%
\and
\thanks{Husheng Li is with the School of Aeronautics and
 Astronautics, Purdue University, West Lafayette, Indiana, USA-47907 (email: husheng@purdue.edu).}
}

\maketitle

\thispagestyle{empty}

\IEEEpeerreviewmaketitle

\begin{abstract}
Integrated sensing and communication (ISAC) is poised to be a defining feature of 6G networks, promising to transform cellular base stations (BSs) into ubiquitous radar sensors. However, a significant gap exists between the theoretical promise of ISAC and the commercial reality of legacy cellular communication infrastructure. Existing communication networks are constrained by fragmented spectrum, blockage-prone environments, and cost-prohibitive high-rate analog-to-digital converters (ADCs). These limitations stifle the high-resolution sensing required for emerging applications. This article advocates a shift from dependence on physical resources to computational synthesis and introduces a unified full stack virtualization framework that upgrades legacy networks with minimal hardware changes, spanning signal generation, propagation, and acquisition. Specifically, we virtualize signal generation via space-time -frequency synthesis across distributed BSs to synthesize a larger effective aperture and a wider effective bandwidth. We then virtualize signal propagation by leveraging environmental multipath and digital maps to reinterpret reflections as massive virtual arrays. Finally, we virtualize signal acquisition using sub Nyquist strategies to bypass sampling bottlenecks. We demonstrate that by trading computation for hardware, commercial networks can achieve fine-grained sensing without expensive retrofitting.
\end{abstract}

\section{Introduction}

The evolution towards 6G networks heralds a paradigm shift from purely connecting devices to perceiving the physical world. Integrated sensing and communication (ISAC)~\cite{ISAC1} has emerged as a cornerstone technology for this next-generation infrastructure. By unifying radar sensing and data transmission within a single hardware and spectral footprint, ISAC transforms cellular base stations (BSs) from simple data deliveries into ubiquitous radio sensors. This integration allows operators to leverage the massive existing deployment of cellular BSs to maximize spectral efficiency and hardware utilization without the need for separate, dedicated radar networks.


\begin{figure*}[!t]
     \centering
 \includegraphics[width=5.5in]{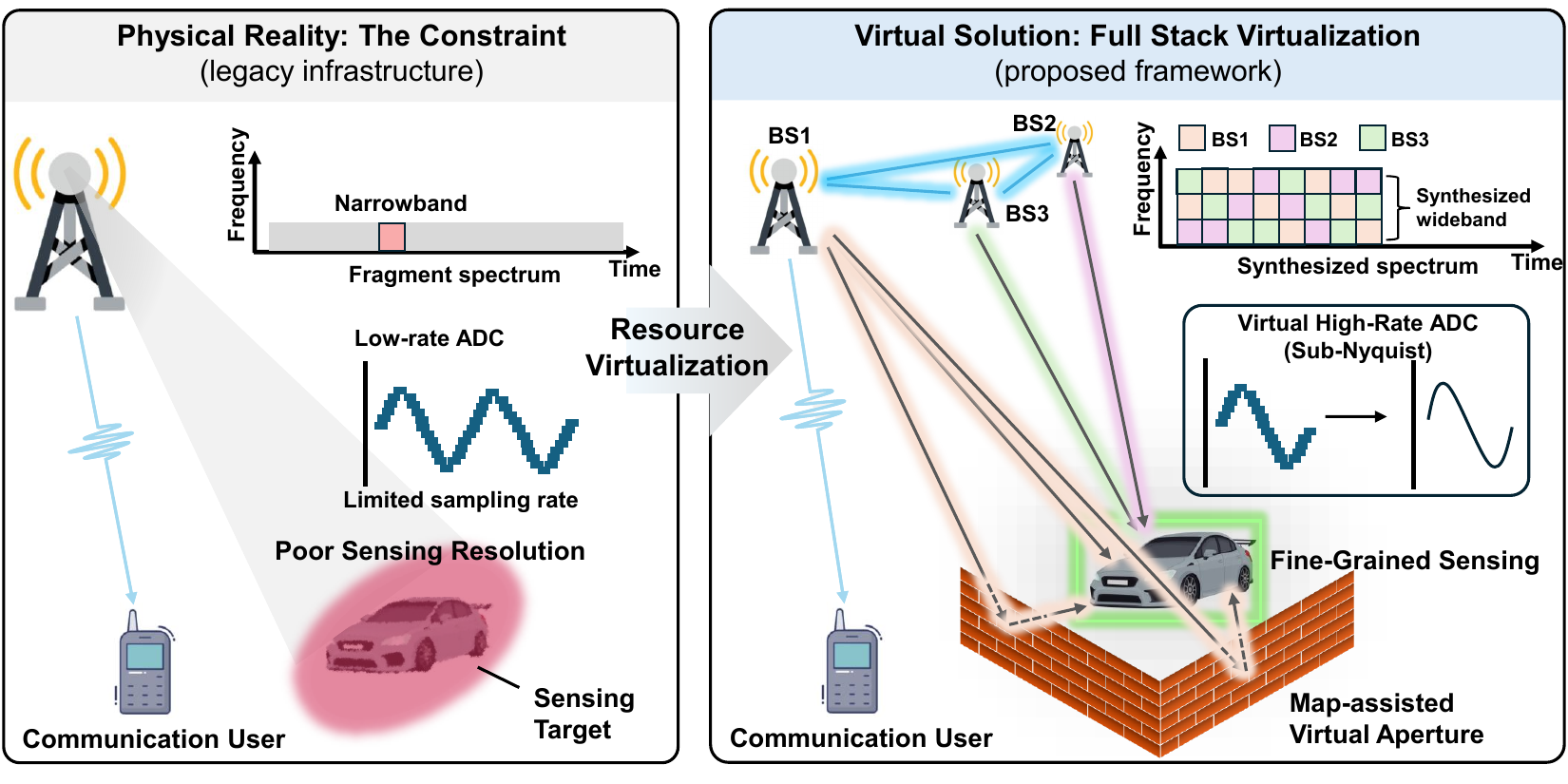}\\
 \caption{Conceptual overview of the proposed full-stack resource virtualization framework. Legacy narrowband, low-rate cellular infrastructure (left) is transformed into a fine-grained ISAC sensing system (right) by virtualizing signal generation at the network level, signal propagation via map-assisted virtual apertures, and signal acquisition through sub-Nyquist virtual high-rate ADCs.}
 \label{fig:overview}
\end{figure*}

This pervasive sensing capability unlocks a vast array of transformative applications~\cite{ISAC_applications}, yet new use cases impose stringent performance demands that go far beyond basic target detection. While coarse sensing might suffice for simple presence monitoring, emerging cyber-physical systems (CPSs) require high-fidelity perception to function safely and effectively. Autonomous vehicles navigating dense urban traffic rely on centimeter-level localization and precise velocity estimation to avoid collisions~\cite{vehicular}. Smart manufacturing facilities depend on the accurate tracking of mobile robots to synchronize complex production lines~\cite{manufacture}. Similarly, immersive technologies such as the metaverse and digital twins require high-resolution environmental mapping to seamlessly blend physical and virtual realities~\cite{DT}. To support these safety-critical and high-precision interactions, the underlying ISAC system must deliver fine-grained sensing resolution that rivals dedicated instrumentation.

However, a significant gap remains between the theoretical promise of ISAC and the practical capabilities of commercial cellular networks. High-resolution radar sensing is governed by stringent physical requirements: fine angular resolution demands large antenna apertures, fine range resolution requires wide contiguous bandwidths, and fine Doppler resolution necessitates long coherent observation times. Unfortunately, legacy communication infrastructure is often deficient in all three dimensions. Commercial BSs are constrained by limited physical sizes, and user equipment (UE) possesses even smaller apertures, severely limiting the beamforming gain for sensing. Furthermore, while fine range resolution requires multi-gigahertz bandwidths, current 5G New Radio (NR) systems typically provide only a few hundred megahertz in fragmented bands~\cite{release19}. Communication signals are also inherently bursty and intermittent, making it difficult to sustain the long coherent durations required for precise velocity estimation.

Beyond these physical and regulation constraints, the cost of processing high-fidelity sensing data presents a major barrier to widespread adoption. In traditional radar systems such as frequency-modulated continuous wave (FMCW), the transmitted waveform is deterministic and known. This allows receivers to utilize "stretch processing" (de-chirping) to downconvert high-bandwidth echoes into low-bandwidth intermediate frequency (IF) signals, drastically reducing the sampling rate requirements. Conversely, communication signals carry unknown, random information. To fully recover this information without distortion, the receiver must strictly adhere to the Nyquist-Shannon theorem~\cite{Nyquist_shannon} relative to the full signal bandwidth. Since the information density for sensing is significantly lower than for communication, typical radars enjoy relaxed sampling constraints that ISAC receivers cannot afford. Consequently, achieving the gigahertz-level bandwidths required for centimeter-level ISAC sensing would demand ultra-high-rate ADCs. These components are prohibitively expensive and generate unmanageable data loads, rendering high-end sensing impractical for cost-sensitive commercial transceivers.

To bridge this gap between theoretical promise and hardware reality, we propose a new design paradigm called “Virtualized Fine-Grained Sensing on a Budget”. We argue that high-precision ISAC does not require a complete hardware overhaul. Instead it can be achieved with a computational approach that synthesizes the necessary physical resources through advanced signal processing with minimum modification to legacy cellular infrastructure. In this article, we present a unified framework for upgrading legacy infrastructure by virtualizing the sensing stack, from signal generation and channel propagation through to signal acquisition:
\begin{itemize}
    \item \textbf{Virtualizing Signal Generation (Network Level)}: We address the limited bandwidth and aperture of individual BSs through space-time-frequency synthesis ISAC network. The system stitches together measurements across distributed BSs, staggered time intervals, and non-contiguous frequency bands. This process creates a large virtual aperture and a wide virtual bandwidth to enable fine-grained sensing on standard communication protocols.
    \item \textbf{Virtualizing Signal Propagation (Channel Level)}: We further address the physical limitations of small antenna apertures through map-assisted environmental synthesis. By fusing multipath time-of-flight (ToF) measurements with known environmental maps, we treat specular reflections as signals from "virtual radars" rather than interference. This creates a massive virtual antenna array that exponentially expands the effective sensing aperture, significantly lowering the localization error bound and enabling fine-grained sensing even in complex environments.
    \item \textbf{Virtualizing Signal Acquisition (Device Level)}: We tackle the ADC bottleneck through Sub-Nyquist sampling scheme. This approach exploits sparse signal structures to recover wideband information from low-rate samples via controlled aliasing. While applicable to various waveforms, we extend this concept to orthogonal time frequency space (OTFS) modulation~\cite{OTFS} to ensure robustness under high mobility. By strategically placing pilots in the delay-Doppler domain, we allow the receiver to recover high-resolution radar parameters using low-cost hardware that samples at a fraction of the Nyquist rate.
\end{itemize}

Fig.~\ref{fig:overview} illustrates the proposed full-stack virtualization framework. By virtualizing resources across the device, channel, and network, we demonstrate that high-precision ISAC is achievable without a physical overhaul.  This framework proves that fine-grained sensing depends not on specialized hardware, but on the computational synthesis of space, time, and frequency. Consequently, these capabilities can be seamlessly layered onto existing cellular infrastructure, bypassing the economic and logistical barriers of hardware retrofitting. In the following sections, the detailed full-stack are
discussed, respectively. Challenges and future research are discussed following that and conclusions are drawn in the final section.

\begin{figure}[!t]
 \centering
 \includegraphics[width=3.2in]{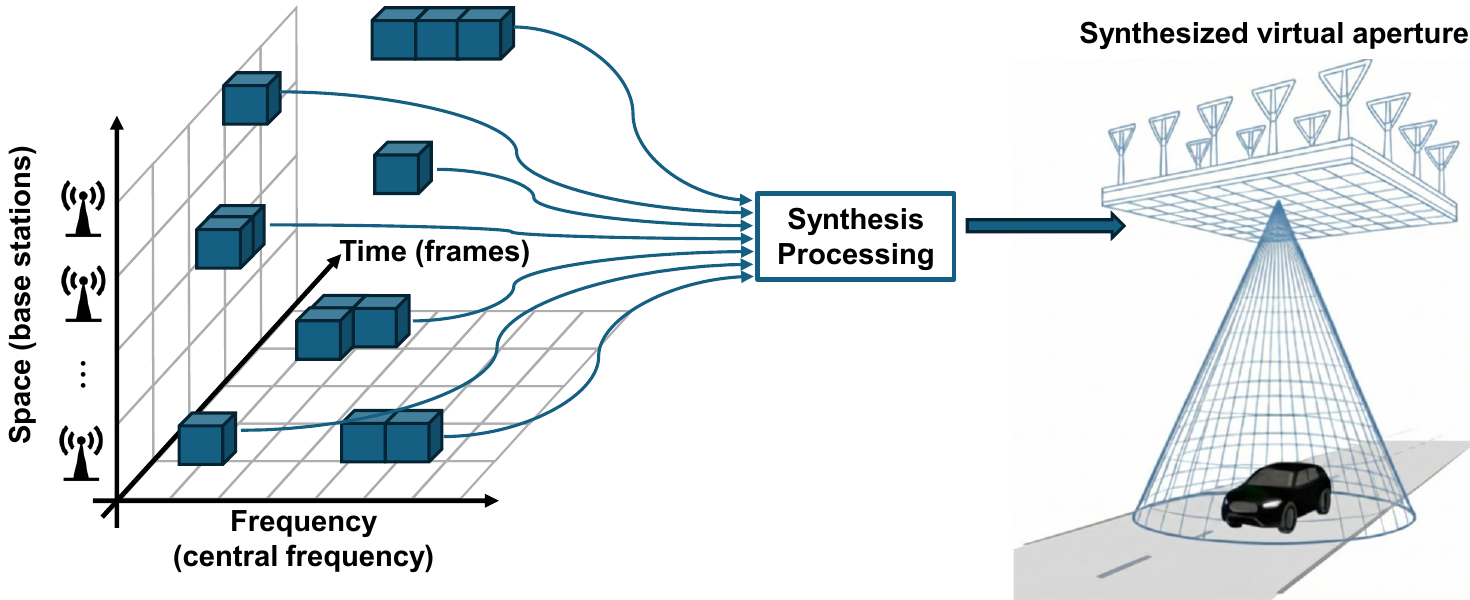}\\
 \caption{Illustration of space-time-frequency synthetic ISAC network.}
 \label{fig:synthetic}
\end{figure}

\section{Virtualizing the Network: From Fragmented Resources to Unified Aperture}\label{sec:space_time_frequency}

The first hurdle in deploying high-precision ISAC is the physical limitation of each transmitter. A commercial BS typically operates alone, constrained by its own bandwidth allocation and fixed antenna size. To approach the resolution of a dedicated radar, we must go beyond what any single site can provide and instead virtualize the network as a whole. Our proposed framework does this by combining resources across space, time, and frequency through network control, without modifying the existing radio front-ends. In particular, multiple legacy BSs, fragmented carriers, and time slots are jointly coordinated to form a space–time–frequency synthetic ISAC network, as illustrated in Fig.~\ref{fig:synthetic}. This virtualization layer effectively enlarges the sensing aperture and bandwidth while keeping the hardware at each individual site unchanged.

\subsection{Frequency Synthesis: Virtual Wideband via Scheduling}
High-resolution ranging is fundamentally limited by signal bandwidth. While dedicated radars sweep gigahertz of contiguous spectrum, 5G spectrum is often fragmented or allocated in narrow slices. Upgrading hardware to support wideband transmission is costly and impractical.

Instead, we exploit the inherent frequency agility of modern cellular networks. We employ a frequency-hopping strategy where the BS switches its carrier frequency across non-contiguous bands over a sequence of time slots. By stitching together these disjoint narrowband measurements, the system constructs a wide virtual bandwidth.
This approach leverages existing 5G NR features such as carrier aggregation and bandwidth parts (BWP). The “synthesis" is purely a scheduling process, compatible with standard schedulers, allowing legacy narrow-band transceivers to emulate wideband radar performance.


\subsection{Time Synthesis: Coherent Integration across Frames}
Fine velocity resolution requires long coherent observation times. However, cellular transmissions are bursty and intermittent. To overcome this, we synthesize a longer coherent processing interval (CPI) by integrating pulses across successive radio frames. This method fits seamlessly within the standard 5G frame structure. By treating periodic synchronization signal blocks (SSBs) or channel state information reference signals (CSI-RS) as "slow-time" radar pulses, we can accumulate Doppler information over hundreds of milliseconds without disrupting user traffic or requiring continuous transmission.

\subsection{Spatial Synthesis: The Distributed Phased Array}

Finally, we synthesize a massive aperture by fusing measurements from multiple BSs. Unlike a monostatic radar that relies on a single reflection path, our network operates as a multistatic system. By combining observations from different viewing angles, the network constructs a virtual aperture that spans kilometers, in a manner analogous to synthetic aperture radar (SAR)~\cite{SAR}, thereby surpassing the diffraction limit of individual small-aperture BSs.

Crucially, to ensure compatibility with legacy backhaul, we adopt a self-coherent but inter-node noncoherent processing model~\cite{synthetic_isac}. Commercial communication networks operate on coarse synchronization standards (e.g., microsecond-level frame alignment) which are sufficient for data transmission but inadequate for sensing, which demands nanosecond-level phase coherence. Since legacy base stations lack the ultra-precise timing hardware (e.g. white rabbit clocks~\cite{white_rabbit}) required to bridge this gap, our model relaxes the requirement: each BS maintains phase coherence only internally. The central processor jointly fuses the measurements rather than the raw phases, allowing the system to run on standard commercial backhaul networks.

\subsection{The Delay-Velocity Coupling Insight}


A key insight from our synthetic ISAC analysis~\cite{synthetic_isac} is that frequency hopping alters the joint identifiability of delay and velocity. Because each pulse uses a different carrier, the received slow time phase contains both a delay dependent carrier phase and a Doppler phase whose sensitivity scales with the carrier frequency. As a result, delay and velocity are no longer cleanly separable and a delay velocity coupling term appears, governed by the hop schedule through the covariance between pulse times and carrier terms.

This coupling has two practical implications. First, it can be managed through scheduling: a balanced hopping pattern that removes this covariance reduces the coupling and yields more stable joint estimation. Second, when bandwidth and time are in competition, our CRLB analysis suggests prioritizing synthesized bandwidth over extending the coherent interval. Widening the synthesized bandwidth strengthens delay information and also boosts Doppler sensitivity through higher carrier levels, whereas extending the observation time provides primarily linear accumulation. This makes spectral agility across fragmented bands especially valuable for legacy networks operating under bursty traffic constraints.


\section{Virtualizing the Environment: Expanding Aperture via Virtual Arrays}\label{sec:map_assisted}


\begin{figure}[!t]
 \centering
 \includegraphics[width=3.2in]{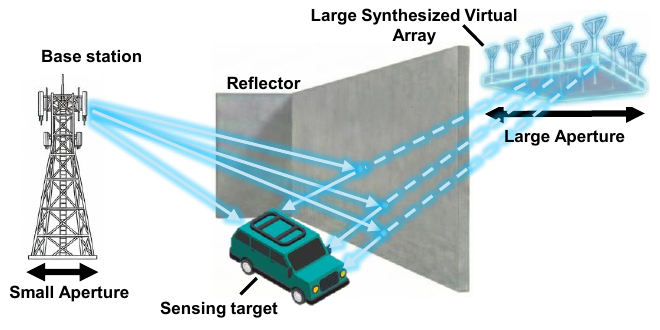}\\
 \caption{Illustration of map-assisted environmental synthesis.}
 \label{fig:map_assited}
\end{figure}

While network synthesis stitches together distributed BSs, the angular resolution of any single BS remains fundamentally limited by its physical antenna size. To achieve fine-grained angular sensing without deploying massive physical arrays, we consider virtualize the propagation channel itself. Rather than treating multipath as a nuisance, we can interpret it as a source of additional virtual sensing points by leveraging digital maps of the environment. By combining map information with basic propagation geometry, the receiver can reinterpret reflections as signals arriving from a synthesized array of virtual sensors, as illustrated in Fig.~\ref{fig:map_assited}. This map-assisted environmental synthesis effectively enlarges the sensing aperture without adding any physical antennas.

\subsection{The Virtual Array Principle}

In rich scattering environments, each dominant specular path can be modeled as originating from a virtual radar mirrored across the reflector. By exploiting these paths, we can effectively expand the sensing aperture from the physical dimensions of the BSs to the geometric dimensions of the surrounding environment. Conceptually, this is also analogous to the principle of SAR: instead of synthesizing a large
aperture through platform motion, we synthesize it through map-aware exploitation of multipath.

The core innovation is map-assisted environmental synthesis~\cite{map_assisted}. By utilizing stepped-frequency synthesis to generate high-precision range profiles and fusing these with a digital map (e.g., building footprints), the system can mathematically invert multipath reflections. This transforms a single physical BS into a cluster of synchronized virtual radars surrounding the target. This massive virtual aperture resolves the multipath ambiguity and dramatically lowers the Cramér-Rao lower bound (CRLB) for localization, enabling centimeter-level accuracy using standard commercial hardware.

\subsection{Addressing mmWave Reflection Loss}
A critical challenge in extending this concept is the behavior of high-frequency signals. As frequencies climb into the millimeter wave (mmWave) and terahertz bands, reflection losses from common building materials (e.g. concrete and glass) increase significantly, potentially attenuating the virtual radar signals below the noise floor. This can be mitigated by increasing processing gain through coherent integration across frames/hops and by jointly combining multiple virtual-sensor returns. Importantly, time synthesis describe in Section~\ref{sec:space_time_frequency} directly supports this goal by accumulating energy over a longer CPI, making weak reflections usable for sensing.

    

To mitigate high path loss without adding hardware, we can also rely on the time synthesis mechanism introduced in the previous section. By stitching together pulses over a longer CPI, the system accumulates signal energy over time. This increased processing gain effectively lifts weak multipath reflections out of the noise, allowing the receiver to detect and utilize high-loss non-line-of-sight (NLOS) paths that would otherwise be invisible to standard communication receivers.

\subsection{Robustness Beyond Line-of-Sight}
Environmental synthesis also improves robustness to blockage. Because sensing can rely on reflected paths, the system need not depend exclusively on line-of-sight (LOS). When LOS is blocked, NLOS virtual sensors can sustain tracking. Moreover, the resulting sensing estimates can feed back into communications: once a user is localized, the BS can perform position-aware beam steering toward an effective reflective surface, reducing exhaustive beam-sweeping latency and improving reliability under blockage.

We implemented an initial indoor NLOS experiment to validate the map-assisted virtual-array localization framework~\cite{NLOS_Pu}, with the setup shown in Fig.~\ref{fig:exp_setting_a}. As shown in Fig.~\ref{fig:exp_result_a}, the map-assited approach can deliver a localization performance with centimeter-level error and also the synthesized-wideband configuration significantly improves localization accuracy compared with a single-channel baseline.

\section{Virtualizing the Receiver: High-Fidelity Sensing on Low-Rate Hardware}

The final bottleneck is the Nyquist constraint. While dedicated FMCW radars use stretch processing to lower sampling rates, communication systems cannot use this shortcut because their signals carry random, unknown data. Consequently, achieving the multi-gigahertz bandwidths required for fine-grained sensing seems to imply expensive and power-hungry ADCs. We address this by virtualizing the acquisition process at the device level.

\begin{figure}[!t]
 \centering
 \includegraphics[width=3.5in]{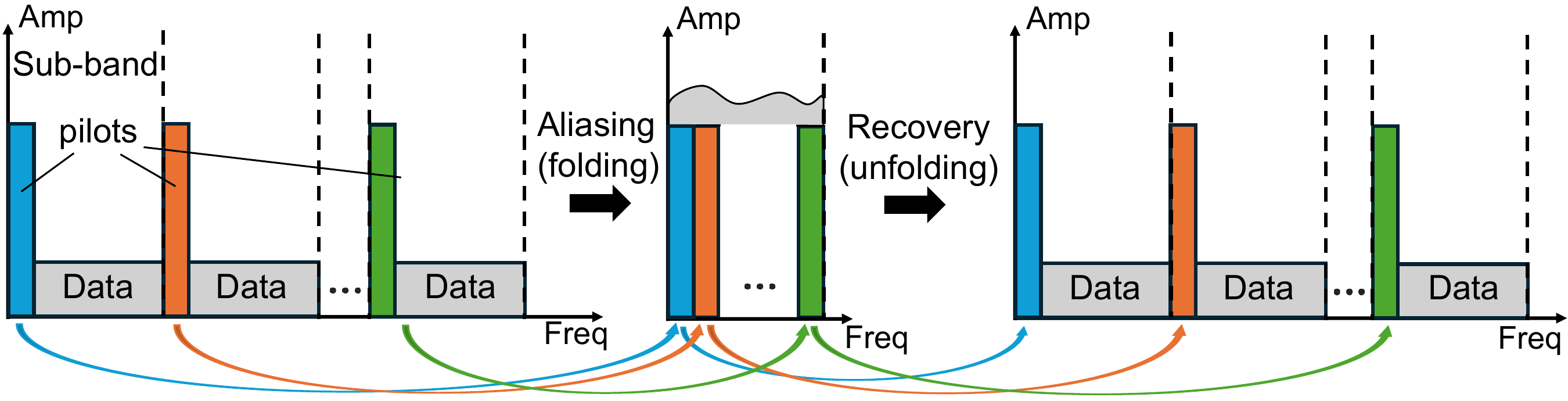}\\
 \caption{Illustration of controlled aliasing under sub-Nyquist sampling, where sparse pilots are designed to fold into distinct regions and are later recovered by dedicated algorithm.}
 \label{fig:aliasing}
\end{figure}

\subsection{Sub-Nyquist Sampling Virtualization}
To break the cost barrier, we introduce an architecture that allows the receiver to violate the Nyquist limit while preserving signal integrity. The core principle is “controlled aliasing”.

When a wideband signal is sampled at a fraction of the Nyquist rate, high-frequency components fold onto lower frequencies. In a standard system, this folding acts as destructive interference that scrambles the information. However, if the transmitted signal utilizes a specific sparse structure, such as isolated pilot tones arranged at calculated intervals, we can predict the exact folding locations of these components. By strategically placing pilots to ensure they alias into orthogonal landing zones in the baseband, the receiver can effectively unfold the signal, as illustrated in Fig.~\ref{fig:aliasing}. This enables the recovery of the full-bandwidth channel response~\cite{OFDM_sub_Nyquist1} or the iterative estimation of sensing parameters and data demodulation using a low-rate, low-cost ADC~\cite{sub_nyquist}.

\subsection{From OFDM to OTFS: Handling Mobility}
While sub-Nyquist strategies have been explored for OFDM radars~\cite{OFDM_sub_Nyquist1}, OFDM struggles in high-mobility scenarios (e.g., high-speed trains or autonomous vehicles) due to severe Doppler shifts.

We extend the sub-Nyquist paradigm to OTFS modulation~\cite{sub_nyquist}. By designing the sensing pilots in the delay-Doppler (DD) domain rather than the time-frequency (TF) domain, we ensure the system remains robust against doubly dispersive channels while maintaining the benefits of reduced sampling rates. Our OTFS-based implementation relies on two mechanisms:
\begin{enumerate}
    \item \textbf{Strategic Pilot Placement:} We embed sensing pilots in specific, isolated bins within the DD domain. We design the placement such that, after passing through the sub-Nyquist sampler, these pilots alias into protected, non-overlapping regions of the spectrum, allowing perfect recovery of the radar parameters.
    
    \item \textbf{Iterative estimation and demodulation:} Under sub-Nyquist sampling, dense data symbols can fold into the sensing measurements. We therefore iterate between joint channel/sensing-parameter estimation and data demodulation/decoding, using the decoded symbols to reconstruct and cancel the data-induced component. This iterative refinement progressively suppresses self-interference, enabling high-resolution sensing while maintaining reliable data demodulation.
\end{enumerate}

\begin{figure}[!t]
   \centering
   \subfigure[Indoor NLOS scenario for validating the map-assisted localization scheme, showing the colocated radar transceiver and sensed object.]
   {
       \label{fig:exp_setting_a}
       \includegraphics[width=0.7\columnwidth]{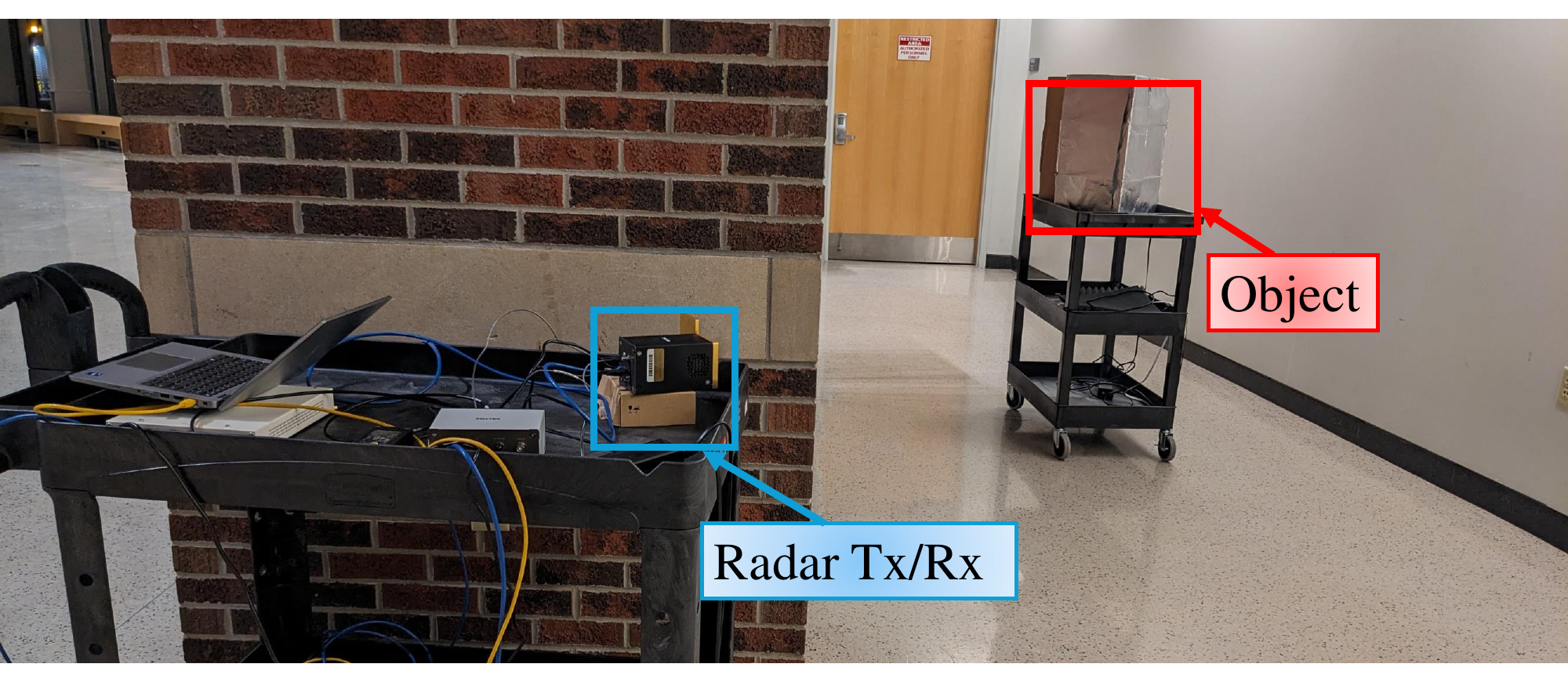}
   }
   \hspace{0.01\linewidth}
   \subfigure[Laboratory setup for validating the sub-Nyquist OTFS-ISAC receiver with two closely spaced targets (Target A and Target B).]
   {
       \label{fig:exp_setting_b}
       \includegraphics[width=0.7\columnwidth]{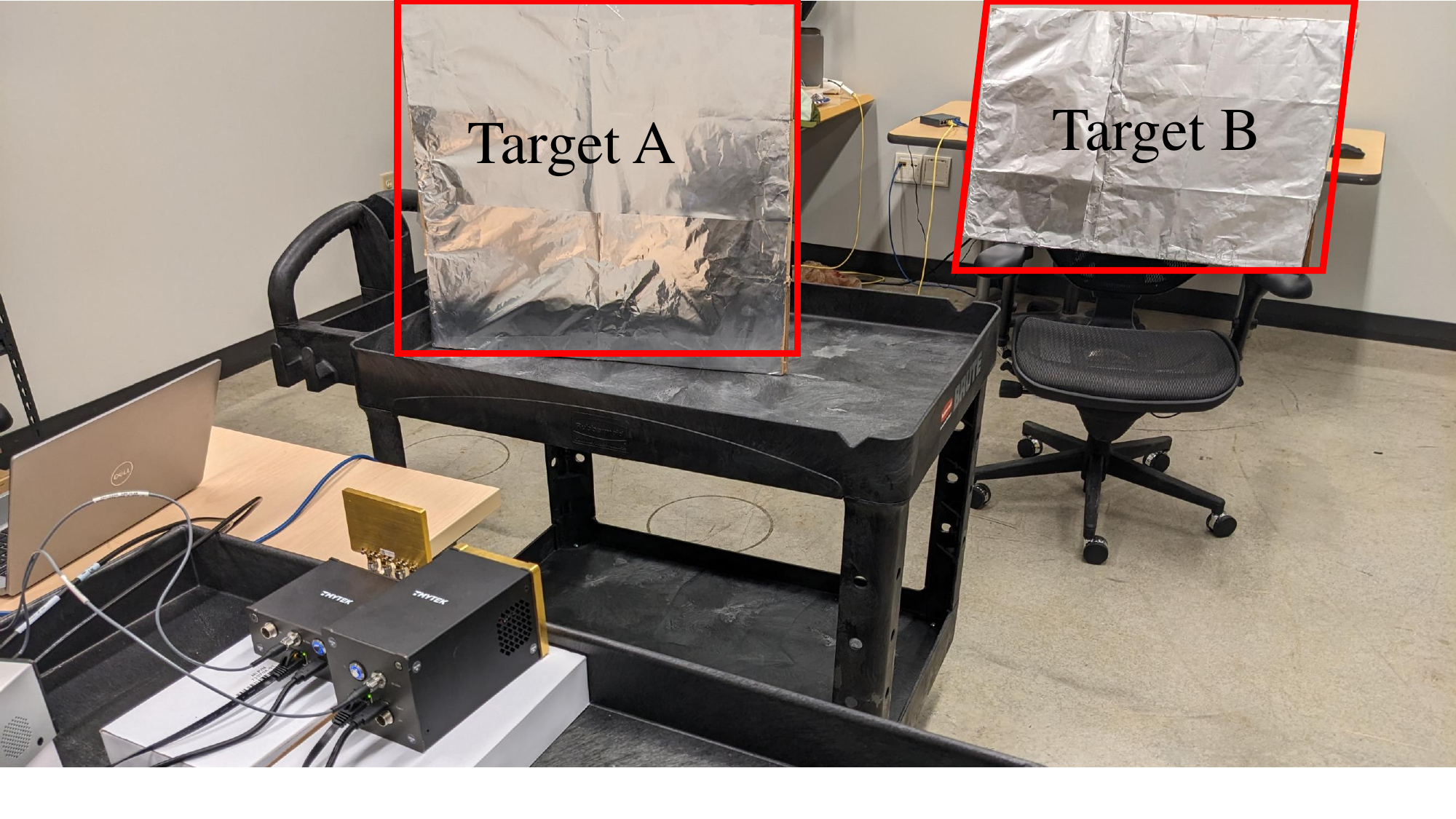}
   }
   \hspace{0.01\linewidth}
   \caption{Experimental setups.}
   \label{fig:exp_setting}
\end{figure}

\begin{figure}[!t]
   \centering
   \subfigure[Experimental localization result with single-channel and synthesized-wideband signals.]
   {
       \label{fig:exp_result_a}
       \raisebox{1.5mm}{\includegraphics[width=0.43\columnwidth]{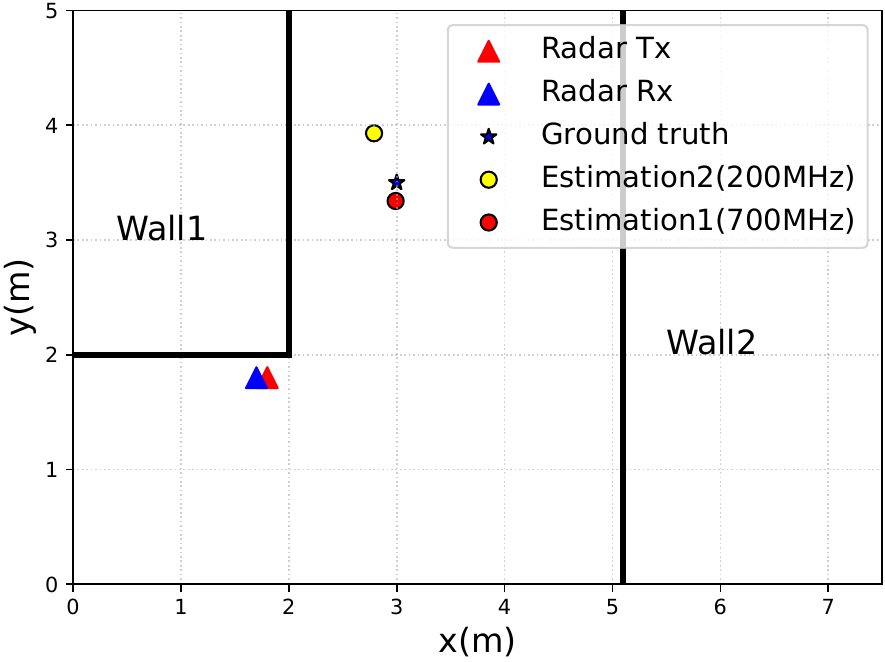}}
   }
   \hspace{0.01\linewidth}
   \subfigure[Experimental range–velocity profile under 16$\times$ sub-Nyquist sampling.]
   {
       \label{fig:exp_result_b}
       \includegraphics[width=0.43\columnwidth]{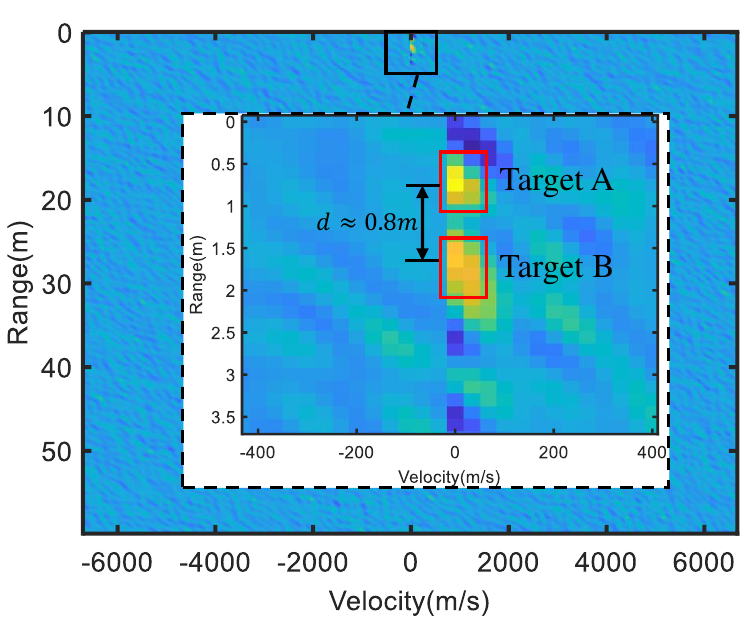}
   }
   \hspace{0.01\linewidth}
   \caption{Experimental results. }
   \label{fig:exp_result}
\end{figure}

To validate the proposed sub-Nyquist OTFS-ISAC receiver, we built a laboratory prototype with two closely spaced targets separated by $0.8\mathrm{m}$, as shown in Fig.~\ref{fig:exp_setting_b}. This distance between the two targets is just beyond the range resolution achievable with a 200~MHz sampling rate. The resulting range–velocity profile under $16\times$ sub-Nyquist sampling (12.5~MHz) is shown in Fig.~\ref{fig:exp_result_b}, where Targets A and B remain clearly resolved despite the reduced ADC rate.

\section{Challenges and Future Directions}

Virtualizing physical resources shifts the complexity burden from hardware to software and network coordination. While this enables high-precision sensing on legacy infrastructure, it introduces specific challenges that define the roadmap for future ISAC research.

\begin{itemize}
\item \textbf{Transitioning to Coherent Distributed Sensing:} While our proposed space-time-frequency framework in Section~\ref{sec:space_time_frequency} effectively utilizes non-coherent fusion to accommodate legacy infrastructure~\cite{synthetic_isac}, this approach trades potential sensing performance gains for feasibility. Achieving a distributed coherent aperture on commercial hardware remains a significant hurdle, as oscillators in low-cost commercial BSs suffer from phase noise and drift, and legacy backhaul lacks nanosecond-level precision.
Except for mandating hardware upgrades like high-stability clocks, future research should prioritize algorithmic phase alignment. Techniques such as “refocusing” (autofocus) algorithms, foundational in SAR imaging, could be adapted to computationally correct phase errors in cellular networks. Furthermore, the network can exploit static environmental clutter as stable “phase anchors”, enabling dynamic self-calibration without the need for dedicated synchronization hardware.

\item \textbf{Environmental Map Dynamics:} The map-assisted multipath-aided approach described in Section~\ref{sec:map_assisted} assumes that the environment is static and the map is accurate. In reality, urban environments are typically dynamic; furniture moves, and “virtual radar” paths can be blocked by temporary obstacles. For the future work, we need to consider to move from using static maps to ISAC-assisted simultaneous localization and mapping (SLAM). The system should not only query the map but actively update it. By continuously verifying the existence of virtual paths, the ISAC system can detect environmental changes in real-time, effectively "healing" the virtual array when physical blockages occur.

\item \textbf{Protocol Standardization:} Current 3GPP standards enforce rigid frame structures optimized for communication, limiting the protocol flexibility needed for arbitrary frequency hopping or specific waveform design. To bridge this gap, future research should focus on adaptive network scheduling that supports flexible sensing numerologies. A potential direction is defining dedicated ``sensing bandwidth parts'' configured to accommodate irregular pilot patterns and non-contiguous resource allocations, enabling these virtualization techniques to coexist seamlessly with standard data traffic.
\end{itemize}

\section{Conclusion}

In this article, we provide an overview of promising approaches for upgrading legacy cellular infrastructure to meet the stringent requirements of fine-grained ISAC sensing with minimal hardware modifications. We introduced a full-stack “resource virtualization” strategy that circumvents physical deficiencies by synthesizing aperture and bandwidth at the network level, exploiting environmental multipath as massive virtual arrays to see around corners, and leveraging sub-Nyquist strategies to bypass the ADC bottleneck at the receiver. This “Virtualized Fine-Grained Sensing on a Budget” paradigm shifts the definition of performance from hardware specifications to computational intelligence, transforming physical constraints into opportunities to ensure that the high-precision “sixth sense” of the network becomes a scalable commercial reality in the 6G era.

\bibliographystyle{IEEEtran}
\bibliography{main}




\end{document}